\documentstyle[12pt,epsf]{article}
\newcommand{\bec}{\begin{center}}
\newcommand{\ec}{\end{center}}

\newcommand{\bee}{\begin{equation}}
\newcommand{\ee}{\end{equation}}
\newfont{\blackboard}{msbm10 scaled\magstep2}
\newcommand{\Z}{\mbox{\blackboard\symbol{"5A}}}
\newcommand{\C}{\mbox{\blackboard\symbol{'103}}}
\begin{document}
\large
\begin{titlepage}
\bec
{\Large\bf  D-Branes and Hilbert Schemes \\}
\vspace*{15mm}
{\bf Yu. Malyuta and T. Obikhod\\}
\vspace*{10mm}
{\it Institute for Nuclear Research\\
National Academy of
Sciences of Ukraine\\
252022 Kiev, Ukraine\\}
{\bf e-mail: interdep@kinr.kiev.ua\\}
\vspace*{35mm}
{\bf Abstract\\}
\ec
The Nakamura's algorithm is applied to
compute the Hilbert scheme for the D-brane
model $\frac{1}{13}(1,2,10)$. 
\end{titlepage}
\section {\bf Introduction}
The moduli space of a D-Brane 
localized at the singular
point of the Calabi-Yau orbifold 
$ \C^{3}/\Z_{n} $
is an interesting object to study, 
because it 
describes phases of models \cite{1.,2.}
\[ \frac{1}{n}(a_{1},a_{2},a_{3}), \]
where characters $ a_{1}, a_{2}, a_{3} $ 
are coprime
to $ n $ and $  \sum_{i=1}^{3}a_{i}=0 $ (mod $ n $).
The D-brane moduli space is isomorphic to the 
Hilbert scheme, which is 
the fine moduli space of $ \Z_{n} $-clusters 
\cite{3.,4.,5.}. Nakamura \cite{3.}
proposes an algorithm to compute Hilbert schemes as 
toric varieties given by fans.\\
\hspace*{6mm}The purpose of the present paper is to
apply the Nakamura's algorithm to compute the 
Hilbert scheme for the model
\[ \frac{1}{13}(1,2,10), \]
which has been studied in \cite{4.} 
by another method.
\section {\bf Constructions}
In this section we give a sketch of  Nakamura's
constructions \cite{3.}. 
\\
\hspace*{6mm}Proposition 1. 
For every $ \Z_{n} $-cluster 
the defining equations can
be written as equations in one 
of the two following forms: either 
\bec
\begin{tabular}{cccc}
$x^{a+d+1}=\lambda y^{b}z^{f}$ &&
$y^{b+1}z^{f+1}=\mu \nu x^{a+d}$ & \\
$y^{b+e+1}=\mu z^{c}x^{d}$ &&
$z^{c+1} x^{d+1}=\lambda \nu y^{b+e}$ &
\hspace*{0.3cm}
$ xyz=\lambda \mu \nu $ 
\hspace*{0.8cm} ($\uparrow$)\\
$z^{c+f+1}=\nu x^{a}y^{e}$ &&
$x^{a+1} y^{e+1}=\lambda \mu z^{c+f}$ & \\
\end{tabular}
\ec
for some $a,b,c,d,e,f\ge0$; or
\bec
\begin{tabular}{cccc}
$ x^{a+d}=\beta \gamma y^{b-1}z^{f-1} $ & &
$ y^b z^f=\alpha x^{a+d-1} $ &  \\
$ y^{b+e}=\alpha \gamma z^{c-1}x^{d-1} $ & & 
$ z^c x^d=\beta y^{b+e-1} $ &  
\hspace*{0.3cm}
$ xyz=\alpha \beta \gamma $ \hspace*{0.7cm} 
($\downarrow$)\\
$ z^{c+f}=\alpha\beta x^{a-1}y^{e-1} $ & & 
$ x^a y^e=\gamma z^{c+f-1} $ &  \\
\end{tabular}
\ec
for some $a,b,c,d,e,f\ge1$.\\
\hspace*{6mm}Proposition 2. The fan of the Hilbert 
scheme is given by the cones with the vertexes
(divisors)
\bec
$ P=(bc+bf+ef+b+c+e+f+1, ac+cd+df+d, ab+ae+de+a) $\\
$ Q=(bc+bf+ef+b, ac+cd+df+a+d+c+f+1, ab+ae+de+e) $\\
$ R=(bc+bf+ef+f, ac+cd+df+c, ab+ae+de+a+b+d+e+1) $\\
\ec
for the equations ($\uparrow$), and by the cones
with the vertexes (divisors)
\bec
$ P=(bc+bf+ef-b-c-e-f+1, ac+cd+df-d, ab+ae+de-a) $\\
$ Q=(bc+bf+ef-b, ac+cd+df-a-d-c-f+1, ab+ae+de-e) $\\
$ R=(bc+bf+ef-f, ac+cd+df-b, ab+ae+de-a-b-d-e+1) $\\
\ec
for the equations ($\downarrow$).
\section {\bf Computations}
In this section we compute the Hilbert scheme
for the model $\frac{1}{13}(1,2,10)$.\\
\hspace*{6mm}Figure 1 shows the McKay quiver 
tesselated by tripods for $\frac{1}{13}(1,2,10)$.
The corresponding monomial representation of 
this quiver is illustrated in Figure 2.\\
\hspace*{6mm}Comparing the hexagons of Figure 1
marked by identical numbers and using the 
corresponding monomials of Figure 2, 
we infer the following 
equations for $\Z_{13}$-clusters:
\bec
\begin{tabular}{lcccclcccclcccccr}
$x=y^{2}z$ &&&&& $y^{4}=z^{2}x$ &&&&& 
$z^{2}=xy^{3}$ &&&&&& (1) \\
$x=y^{2}z$ &&&&& $y^{5}=z$ &&&&& 
$z^{3}=\hspace*{2.3mm}y^{2}$  &&&&&& (2) \\
$x^{2}=y$  &&&&& $y^{4}=z^{2}x$ &&&&& 
$z^{3}=\hspace*{2.3mm} y^{2}$  &&&&&&(3) \\
$x^{2}=y$  &&&&& $y^{5}=z$    &&&&& 
$z^{2}=xy^{3}$ &&&&&&(4) \\   
$x^{2}=y$  &&&&& $y^{2}=z^{3}$  &&&&& 
$z^{4}=x$      &&&&&&(5) \\ 
$x^{2}=y$  &&&&& $y^{7}=\hspace*{3.6mm} x$
      &&&&& $z =\hspace*{3.7mm} y^{5}$ 
    &&&&&&(6) \\ 
$x=\hspace*{2.8mm} z^{4}$  &&&&& 
$y^{2}=z^{3}$  &&&&& $z^{8}=\hspace*{2mm} y$ 
     &&&&&&(7) \\  
$x^{4}=\hspace*{2mm} z^{3}$ &&&&& 
$y=\hspace*{3.9mm} x^{2}$     &&&&& 
$z^{4}=x$      &&&&&&(8) \\
$x^{7}=\hspace*{2mm} z^{2}$ &&&&& 
$y=\hspace*{3.9mm} x^{2}$     &&&&& 
$z^{3}=x^{4}$  &&&&&&(9) \\ 
$x^{10}=\hspace*{1.8mm} z$ &&&&& 
$y=\hspace*{3.9mm} x^{2}$   &&&&& 
$z^{2}=x^{7}$  &&&&&&(10) \\
$x^{13}=1$ &&&&& $y=\hspace*{3.9mm} x^{2}$
     &&&&& $z=x^{10}$     &&&&&&(11) \\  
$x=y^{7}$ &&&&& $y^{13}=1$     &&&&& 
$z=\hspace*{2mm} y^{5}$     &&&&&&(12) \\ 
$x=\hspace*{2.8mm} z^{4}$  &&&&& $y=z^{8}$
      &&&&& $z^{13}=1$     &&&&&&(13) \\
\end{tabular}\\
\ec
Note that we write this equations
omiting multiplicative constants $\lambda, 
\mu, \nu, \alpha, \beta,\gamma $.
Note also that we require the 
agreement of our equations
with the defining equations of Proposition 1.
This requirement gives us the exponents of
$x^{*}, y^{*}, z^{*} $:
\bec
\begin{tabular}{llllllccl}
$a=1$, & $b=2$, & $c=2$, & 
$d=1$, & $e=3$, & $f=1$ &&& for (1) \\
$a=0$, & $b=2$, & $c=1$, & 
$d=0$, & $e=2$, & $f=1$ &&& for (2) \\
$a=0$, & $b=1$, & $c=2$, & 
$d=1$, & $e=2$, & $f=0$ &&& for (3) \\
$a=1$, & $b=1$, & $c=1$, & 
$d=0$, & $e=3$, & $f=0$ &&& for (4) \\
$a=1$, & $b=1$, & $c=3$, & 
$d=0$, & $e=0$, & $f=0$ &&& for (5) \\
$a=0$, & $b=1$, & $c=0$, & 
$d=1$, & $e=5$, & $f=0$ &&& for (6) \\
$a=0$, & $b=0$, & $c=3$, & 
$d=0$, & $e=1$, & $f=4$ &&& for (7) \\
$a=1$, & $b=0$, & $c=0$, & 
$d=2$, & $e=0$, & $f=3$ &&& for (8) \\
$a=4$, & $b=0$, & $c=0$, & 
$d=2$, & $e=0$, & $f=2$ &&& for (9) \\
$a=7$, & $b=0$, & $c=0$, & 
$d=2$, & $e=0$, & $f=1$ &&& for (10) \\
$a=10$, & $b=0$, & $c=0$, & 
$d=2$, & $e=0$, & $f=0$ &&& for (11) \\
\end{tabular}\\
\begin{tabular}{llllllccl}
$a=0$, & $b=7$, & $c=0$, & 
$d=0$, & $e=5$, & $f=0$ &&& for (12) \\
$a=0$, & $b=0$, & $c=8$, & 
$d=0$,& $e=0$, & $f=4$ &&& for (13) \\
\end{tabular}\\
\ec
\hspace*{6mm}Using these exponents and 
Proposition 2, we obtain 
the following cones of the fan: 
\bec
\begin{tabular}{lclclcl}
$P=(2, 4, 7)$  & & $Q=(7, 1, 5)$
  & & $R=(8, 3, 2)$ && for (1) \\
$P=(13, 0, 0)$  & & $Q=(8, 3, 2)$
  & & $R=(7, 1, 5)$ && for (2) \\
$P=(8, 3, 2)$  & & $Q=(3, 6, 4)$  & & 
$R=(2, 4, 7)$ &&  for (3) \\
$P=(7, 1, 5)$  & & $Q=(2, 4, 7)$  & & 
$R=(1, 2, 10)$ && for (4) \\
$P=(8, 3, 2)$ & &  $Q=(4, 8, 1)$ & &  
$R=(3, 6, 4)$ && for (5) \\
$P=(7, 1, 5)$ &  & $Q=(1, 2, 10)$ 
&  & $R=(0, 0, 13)$ && for (6) \\
$P=(13, 0, 0)$  & & $Q=(4, 8, 1)$  
& & $R=(8, 3, 2)$ && for (7) \\
$P=(4, 8, 1)$ &  & $Q=(0, 13, 0)$ &  & 
$R=(3, 6, 4)$ && for (8) \\
$P=(3, 6, 4)$ &  & $Q=(0, 13, 0)$  & & 
$R=(2, 4, 7)$ && for (9) \\
$P=(2, 4, 7)$ &  & $Q=(0, 13, 0)$  & & 
$R=(1, 2, 10)$ && for (10) \\
$P=(1, 2, 10)$ &  & $Q=(0, 13, 0)$  & & 
$R=(0, 0, 13)$ && for (11) \\
$P=(13, 0, 0)$ &  & $Q=(7, 1, 5)$  & & 
$R=(0, 0, 13)$ && for (12) \\
$P=(13, 0, 0)$ &  & $Q=(0, 13, 0)$  & & 
$R=(4, 8, 1)$ && for (13) \\
\end{tabular}
\ec
\hspace*{6mm}The Hilbert scheme
for $\frac{1}{13}(1,2,10)$
is illustrated in Figure 3.
\vspace*{15mm}\\
{\bf Acknowledgement}
\vspace*{5mm}\\
The authors would like to thank 
Professor I. Nakamura
for sending his preprint.
\begin{center}
\begin{tabular}{c}
\vspace*{-2.5cm}\\
\hspace*{0.9cm} 
\epsfxsize=21cm\epsffile{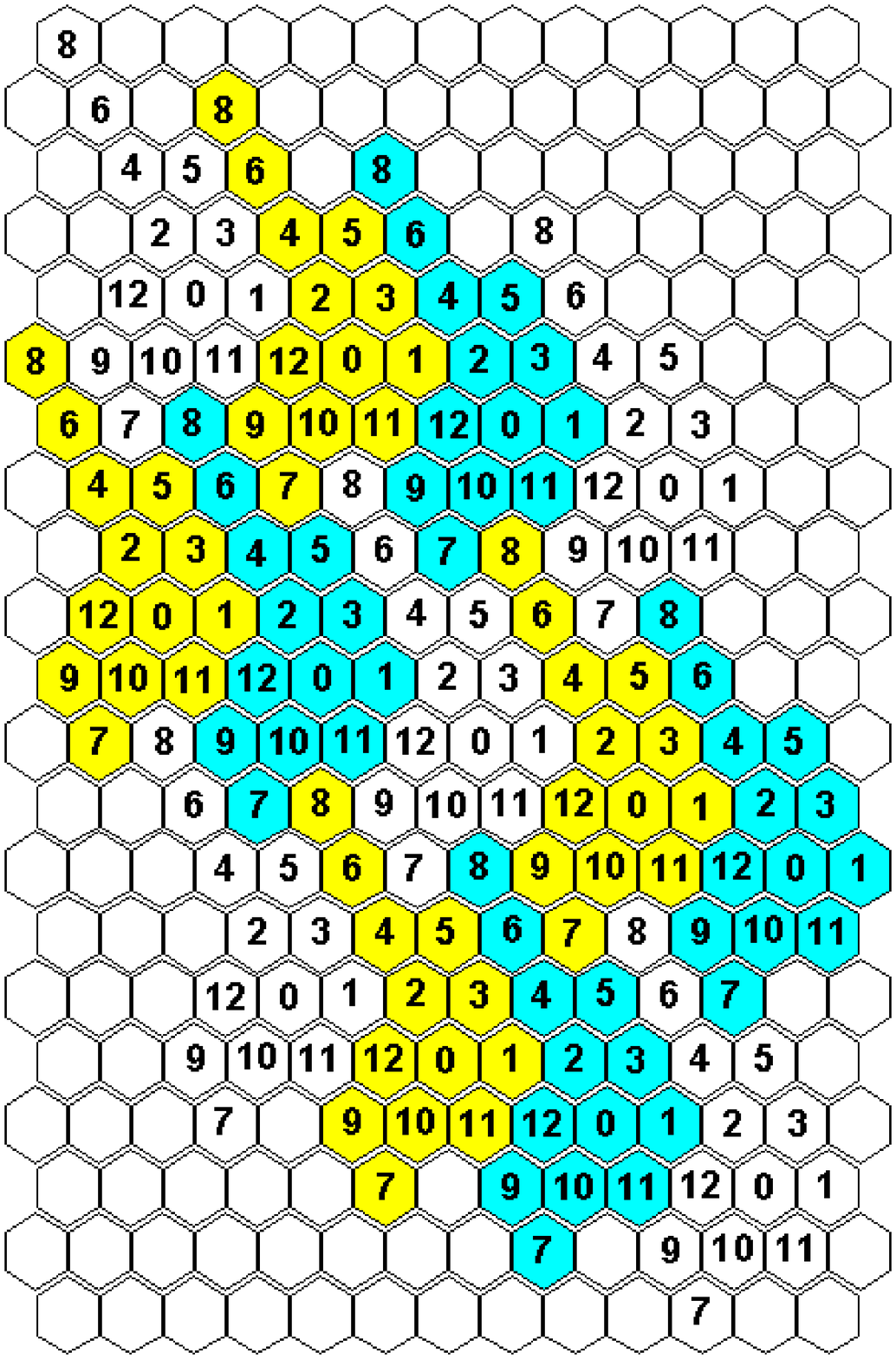}
\vspace*{-2.1cm}\\
\hspace*{-9.1cm}Figure 1\\
\end{tabular}
\end{center}
\newpage
\begin{center}
\begin{tabular}{c}
\vspace*{-2.5cm}\\
\hspace*{0.9cm}
\epsfxsize=21cm\epsffile{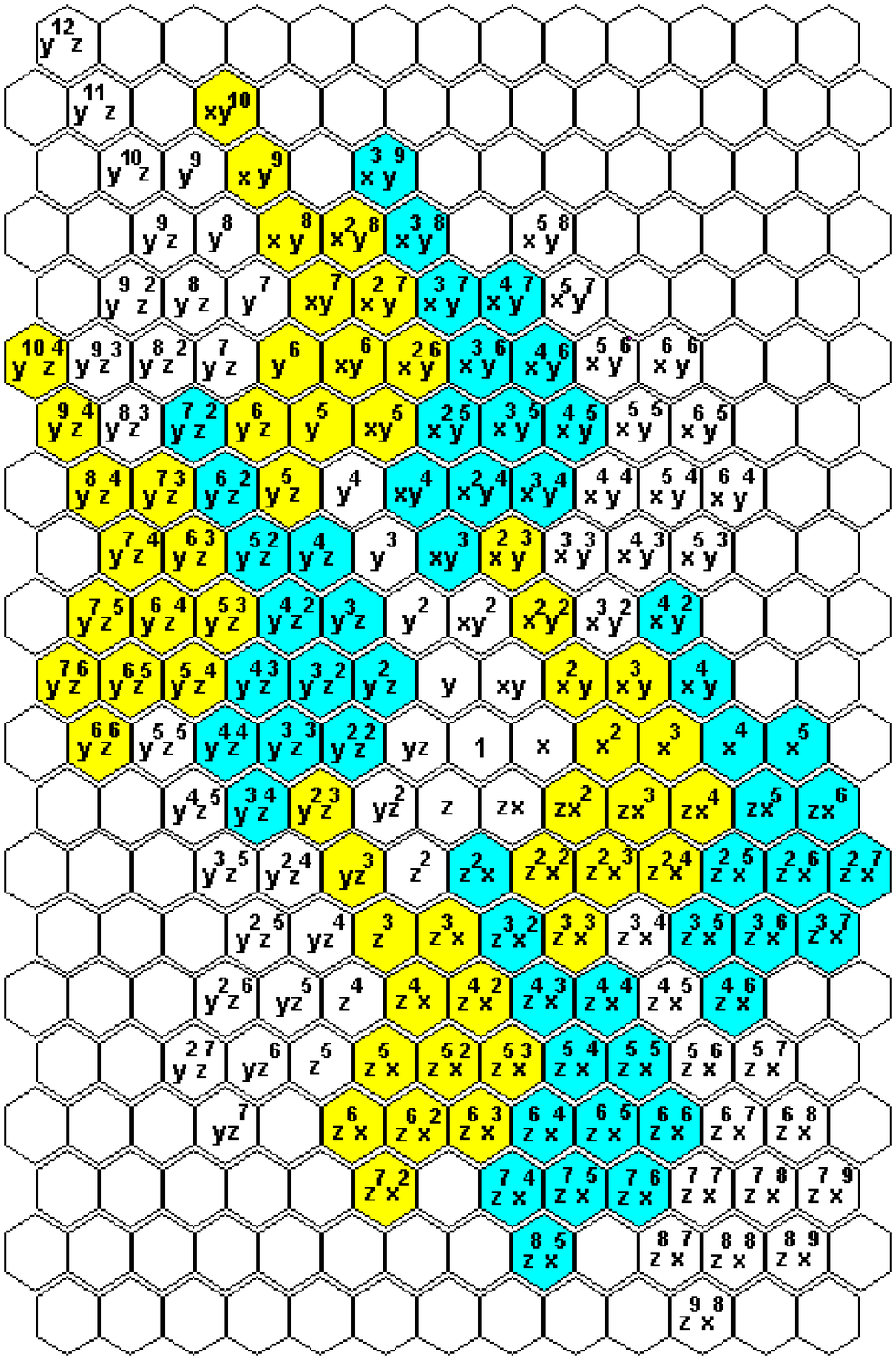}
\vspace*{-2.1cm}\\
\hspace*{-9.1cm}Figure 2\\
\end{tabular}
\end{center}
\newpage
\begin{center}
\begin{tabular}{c}
\vspace*{-2cm}\\
\hspace*{-1.1cm}
\epsfxsize=21cm\epsffile{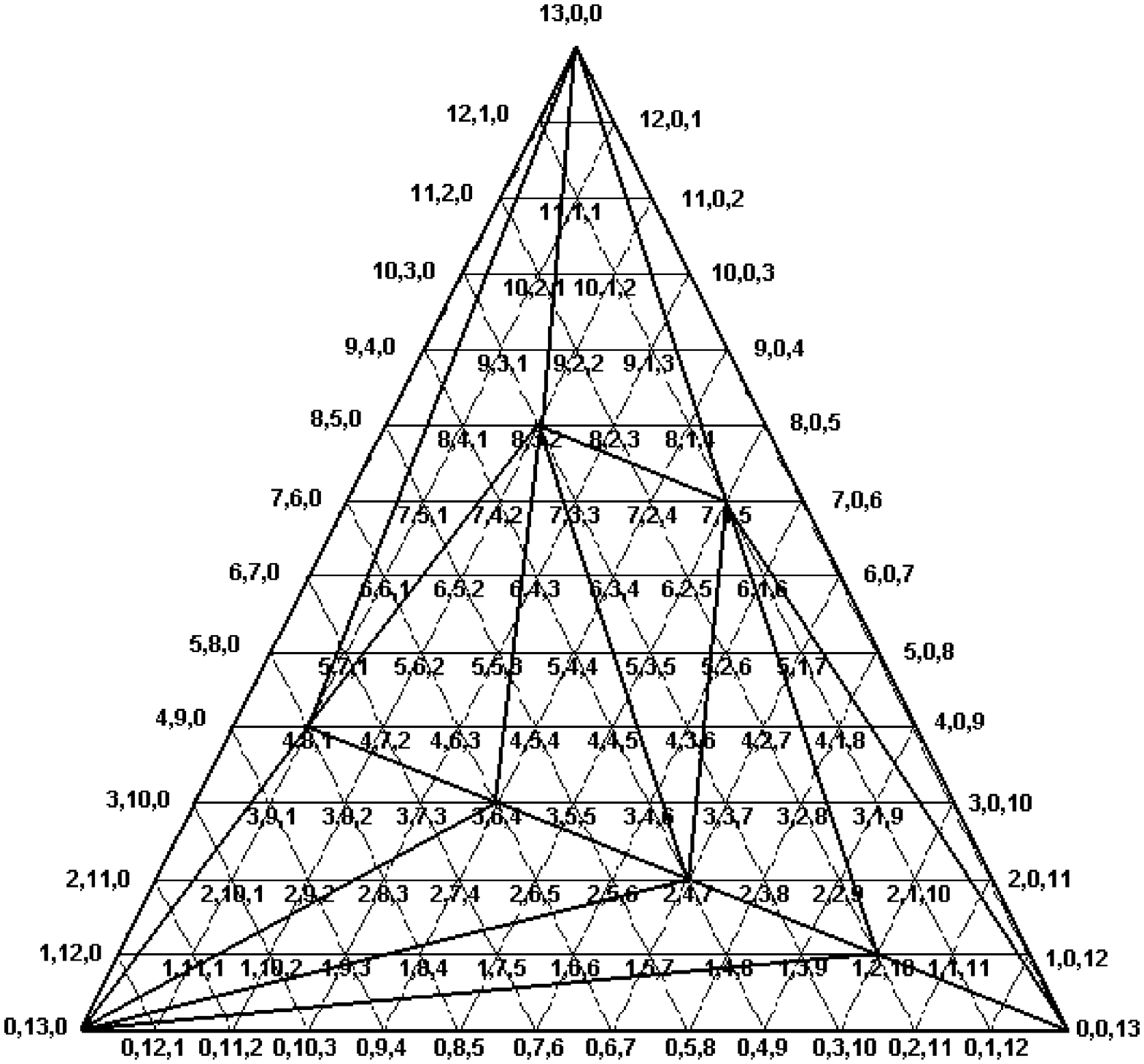}
\vspace*{-3.2cm}\\
\hspace*{0.62cm}
\hspace*{-7.4cm}Figure 3\\
\end{tabular}
\end{center}

\end{document}